\documentclass[preprint,12pt, a4paper]{elsarticle}

\usepackage{amssymb}

\usepackage{lineno}

\usepackage{float}
\usepackage{hyperref}
\usepackage{booktabs}
\usepackage{multirow}
\usepackage{array}
\usepackage[table]{xcolor} 

\usepackage[printonlyused,withpage]{acronym}

\restylefloat{table}

\journal{SoftwareX}

\begin{document}

\acrodef{LIGO}{Laser Interferometer Gravitational Wave Observatory}
\acrodef{SNR}{Signal to Noise Ratio}
\acrodef{aLIGO}{Advanced LIGO}
\acrodef{AEI}{Albert Einstein Institute}
\acrodef{CBC}{Compact Binary Coalescence}
\acrodef{CDS}{Control and Data System}
\acrodef{CIS}{Channel Information System}
\acrodef{CIT}{Caltech}
\acrodef{DAQ}{Data Acquisition System}
\acrodef{DMP}{LIGO Data Management Plan}
\acrodef{DMT}{Data Monitoring Toolbox}
\acrodef{DQSEGDB}{Data Quality Segment Database}
\acrodef{E@H}{Einstein at Home}
\acrodef{ECSS}{Extended Collaborative Support Service}
\acrodef{EM}{Electro-magnetic}
\acrodef{ER}{Engineering Run}
\acrodef{FTE}{Full-Time Equivalent}
\acrodef{GCN}{Gamma-ray Coordinates Network}
\acrodef{GLUE}{Grid LSC User Environment}
\acrodef{KAGRA}{Kamioka Gravitational wave detector}
\acrodef{GraCEDb}{Gravitational-wave Candidate Event Database}
\acrodef{GRINCH}{Gravitational-wave Candidate Event Handlers}
\acrodef{GSL}{GNU Scientific Library}
\acrodef{GW}{gravitational wave}
\acrodef{HTC}{High Throughput Computing}
\acrodef{IFO}{interferometer}
\acrodef{ICTS}{International Centre for Theoretical Sciences}
\acrodef{IdP}{Identity Provider}
\acrodef{iLIGO}{Initial LIGO}
\acrodef{IUCAA}{Inter-University Centre for Astronomy and Astrophysics}
\acrodef{KDC}{Key Distribution Center}
\acrodef{LALSuite}{LSC Algorithm Libraries}
\acrodef{LIAM}{LIGO Identity and Access Management}
\acrodef{LDR}{LIGO Data Replicator}
\acrodef{LDG}{LIGO Data Grid}
\acrodef{GEO}{GEO600}
\acrodef{LIGOdv}{LIGO Data Viewer}
\acrodef{LIGOdv-web}{LIGO Data Viewer Web Service}
\acrodef{LOSC}{LIGO Open Science Center}
\acrodef{LSC}{LIGO Scientific Collaboration}
\acrodef{LSCSoft}{LSC Software Repositories}
\acrodef{LSST}{Large Synoptic Survey Telescope}
\acrodef{LVAlert}{LIGO-Virgo Alert System}
\acrodef{LVC}{\ac{LSC} and the Virgo Collaboration}
\acrodef{LVCN}{LIGO Virgo Computing Network}
\acrodef{NDS}{Network Data Server}
\acrodef{O1}{Observing Run 1}
\acrodef{O2}{Observing Run 2}
\acrodef{O3}{Observing Run 3}
\acrodef{O4}{Observing Run 4}
\acrodef{sO1}{First Observing Run}
\acrodef{sO2}{Second Observing Run}
\acrodef{ODC}{Online Detector Characterization}
\acrodef{OSG}{Open Science Grid}
\acrodef{PE}{Parameter Estimation} 
\acrodef{RDS}{Reduced Data Set}
\acrodef{SDSC}{San Diego Supercomputer Center}
\acrodef{SP}{Service Provider}
\acrodef{TACC}{Texas Advanced Computing Center}
\acrodef{XSEDE}{Extreme Science and Engineering Discovery Environment}
\acrodef{VDT}{Virtual Data Toolkit}
\acrodef{EO}{Engineering and Operations}
\acrodef{COS}{Collaboration Operations Support}
\acrodef{ABB}{Application Building Blocks}
\acrodef{DHS}{Data Handling Services}
\acrodef{DetChar}{Detector Characterization}
\acrodef{GDS}{Global Diagnostics System}
\acrodef{DA}{Data Analysis}
\acrodef{DQ}{Data Quality}
\acrodef{SWIG}{the Simple Wrapper and Interface Generator}
\acrodef{GRB}{gamma-ray burst}
\acrodef{FRB}{fast radio burst}
\acrodef{LIGO}{Laser Interferometer Gravitational-Wave Observatory}
\acrodef{ASKAP}{Australian Square Kilometre Array Pathfinder}
\acrodef{CHIME}{Canadian Hydrogen Intensity Mapping Experiment}
\acrodef{DetChar}{Detector Characterization}
\acrodef{BBH}{binary black hole}
\acrodef{BNS}{binary neutron star}
\acrodef{NSBH}{neutron star - black hole}
\acrodef{EM}{electromagnetic}
\acrodef{GCN}{Gamma-ray Coordination Network}
\acrodef{Fermi}{Fermi Gamma-ray Burst Monitor}
\acrodef{SWIFT}{The Neil Gehrels Swift Observatory}
\acrodef{PCSE}{Department of Physics, Computer Science and Engineering}
\acrodef{CNU}{Christopher Newport University}
\acrodef{IPN}{Interplanetary Gamma-Ray Burst Timing Network}
\acrodef{MOU}{Memorandum of Understanding}
\acrodef{DQSEGDB}{Data Quality Segment Database}

\begin{frontmatter}



\title{DQSEGDB:  A time-interval database for storing gravitational wave observatory metadata}


\author{R.~P.~Fisher$^{1}$,  G.~Hemming$^{2}$, M.~A.~Bizouard$^{3}$, D.~A.~Brown$^{4}$, P.~F.~Couvares$^{5}$,  F.~Robinet$^{6}$, D.~Verkindt$^{7}$} 

\address{$^{1}$CNU, 1 Avenue of the Arts, Newport News, VA 23606, USA, $^{2}$European Gravitational Observatory (EGO), I-56021 Cascina, Pisa, Italy, $^{3}$Artemis, Universit\'e C\^ote d'Azur, Observatoire C\^ote d'Azur, CNRS, CS 34229, F-06304 Nice Cedex 4, France, $^{4}$Syracuse University, Syracuse, NY 13244, USA, $^{5}$LIGO, California Institute of Technology, Pasadena, CA 91125, USA, $^{6}$LAL, Univ. Paris-Sud, CNRS/IN2P3, Universit\'e Paris-Saclay, F-91898 Orsay, France, $^{7}$Laboratoire d'Annecy de Physique des Particules (LAPP), Univ. Grenoble Alpes, Universit\'e Savoie Mont Blanc, CNRS/IN2P3, F-74941 Annecy, France}

\begin{abstract}
The \ac{DQSEGDB} software is a database service, backend API, frontend graphical web interface, and client package used by the \ac{LIGO}, Virgo, GEO600 and the Kamioka  Gravitational  wave  detector for storing and accessing metadata describing the status of their detectors. The \ac{DQSEGDB} has been used in the analysis of all published detections of gravitational waves in the advanced detector era. The DQSEGDB currently stores roughly 600 million metadata entries and responds to roughly 600,000 queries per day with an average response time of 0.223 ms. 
\end{abstract}


\end{frontmatter}

\section{Motivation and significance}
\label{motivation}
\acresetall
Gravitational Waves are disturbances in the metric of space-time that propagate through the universe and carry information about the astrophysics of sources that generate them.  For current detectors, these sources must be massive objects moving with high accelerations \cite{300yrs}.\acused{GW} Although \acp{GW} may have very large amplitudes at their origin, they typically travel extra-galactic distances to reach the Earth.  Because gravity couples weakly to matter, \acp{GW} are difficult to generate and detect.  When these waves reach the Earth, their strength is such that their resulting spacetime perturbation changes measurements of length by 1 part in $10^{20}$, equivalent to changing a distance of 1 AU by the width of 1 atom.  This results in an extremely small signal even if detected with kilometer-scale detectors \cite{creightonbook2011}. The mission of the \ac{LVC} is to detect these weak signals in order to advance our understanding of the universe. Since 2015, \ac{GW} detections have shown that Einstein's theory of general relativity holds for colliding black holes and neutron stars \cite{Abbott:2016blz,gw170817,Abbott:2018mvr}. These discoveries allow us to estimate the number of \ac{BBH} and \ac{BNS} systems in our local universe \cite{Abbott:2016ymx, Abbott:2016drs, Abbott:2016nhf}, and have demonstrated that some \ac{GRB} events are powered by the coalescence of neutron stars \cite{gw170817, Monitor:2017mdv}.

The detection of these \ac{GW}s has been made possible through the colossal effort of thousands of scientists to develop a tremendously precise set of \acp{IFO} and an ecosystem of computational infrastructure that enables the capture and analysis of the data generated by these instruments. The \ac{DQSEGDB} occupies one critical space in this infrastructure.  
The \ac{DA} algorithms require information about the state of the \ac{IFO}s to analyze the observatory data.  This requires the definition and distribution of metadata about the observatory data, which we call \ac{DQ} flags.  

A \ac{DQ} flag is the name given to a set of metadata that describes some portion of the global status of the detector, operation of the instrument, or quality of the data that may impact its analysis.  These flags are critical to the data analyses because a category of them mark the times when the \ac{IFO}s are operating in an optimal state, thereby indicating which observatory data should be analyzed.  Additional flags indicate data that should explicitly not be analyzed, such as when hardware injections are ongoing or when electronics faults cause noise in the \ac{GW} detection channel.  These DQ flags are also called DQ vetoes because they can be used to exclude data from being analyzed \cite{Abbott:2017lwt}.  

The set of data associated with each flag name is the list of times when the state of that flag was known and the list of times when that state was active or inactive, which are compliments within the set of known times. The time periods are contained in a data product known as ``segments", where a segment is a continuous range of time expressed as a half-open GPS time interval [tstart,tend). Within the GW community, the terms \ac{DQ} segments and \ac{DQ} flags are often used interchangeably because of this tight relationship.  Each flag has a unique name. The flag names are associated with their \ac{IFO} identifiers, and are combined in the format [IFO]:[FLAG-NAME].  

\subsection{Initial Detector Databases}
The \ac{DQSEGDB} service and client software were built to replace the aging predecessor services that served the \ac{LIGO} and Virgo collaborations separately during their initial observing runs. These previous services were each able to store hundreds of flags and approximately 2 million individual \ac{DQ} segment and metadata entries by the end of the final science runs of initial \ac{LIGO} and Virgo. The \ac{LIGO} service had became very slow, and would often take 10 to 30 minutes to respond to queries made by GW data analyses. This severely restricted the usability of the service and indicated a strong need for a replacement for advanced \ac{GW} detectors, where the number of flags and number of segments the databases would need to store would grow by factors of hundreds.  An increasing number of new software systems were also unable to use the DQ metadata effectively due to the slow response times of the server. Finally, new user requirements pushed for a redesign of the API and database schema.  These issues led to the \ac{LVC} making the decision to pool their resources and to design a new segment database infrastructure. This led to the development of the \ac{DQSEGDB} software.  


\section{\ac{LVC} Data Landscape and Terminology}
Each \ac{IFO} produces one primary data channel, which contains the measurement of the GW strain, and approximately 200,000 channels of auxiliary data that are used to monitor the status of all the hardware and software components used to produce the primary data.  This data set constitutes approximately 2 TB per day per \ac{IFO}. Customized scripts are used to reduce this huge amount of auxiliary data into approximately 1000 DQ flags per \ac{IFO}.  At the location of the \ac{IFO}s, a set of real-time processes automatically generate segments for a portion of the total \ac{DQ} flags. These processes encode the metadata in XML files that each contain information about the status of these flags for 16 seconds of data. Each of these XML files is about 78 kB in size, which translates to roughly 420 MB of metadata generated per day per IFO. This XML is then transferred via rsync from each \ac{IFO} to the \ac{DQSEGDB} server, which is hosted at the LIGO Laboratory at California Institute of Technology. The \ac{DQSEGDB} server then executes all of the code needed to extract the metadata from the XML files, publishes it to the database, and archives the raw XML files.   

\section{Software Development and Description}
\label{development}


\subsection{Software Design and Architecture}
\label{design_and_performance}
In addition to the requirements that the \ac{DQSEGDB} service be able to respond rapidly while storing a large amount of metadata, several other design requirements and elements of design philosophy were also met when the new software was written.  The database was required to contain both the DQ segments and enough additional metadata to allow the tracking of their provenance.  The service was required to allow remote clients to connect via command line or web GUI, and was to provide the metadata within 15 minutes of its generation.  The API was chosen to provide a RESTful set of URIs with a Resource-Oriented Architecture, compatible with multiple programming languages, and restrictive such that data could not be removed from the database.  A JSON format was chosen for the returned data, which included an option for all provenance metadata.  Additional functionalities were deferred to the client layer to ensure speed at the server.

\subsection{Software Functionalities}
\label{functionalities}

These design requirements led to the current \ac{DQSEGDB} software design. The service as a whole is split into three major components. The first is the primary database server, which is generally labelled the \ac{DQSEGDB}.   
The second is the client software package, which contains both command line tools and a Python package that can be used to query the database.  This set of tools provides many functions as requested by \ac{LVC} scientists, while also satisfying the design requirements listed. The final component is a graphical web interface to the database, which provides a GUI interface that allows collaboration scientists to rapidly access the metadata without needing to write any code.  

The \ac{DQSEGDB} server consists of an Apache layer that calls a custom Python application via the Apache WSGI module.  The Python application uses ODBC to communicate with a MariaDB database instance that hosts a dedicated, normalised database schema that has been designed to optimally host the \ac{DQ} flags, their associated segments, the associated metadata about those segments, and some overall metadata about the data.  The API provides access to all data associated with a given DQ flag through RESTful URIs, formatted as /dq/{IFO}/{FLAG}/{VERSION}.  Within this URI, the data can be downselected based on the information and time interval of interest, using options such as /dq/{IFO}/{FLAG}/{VERSION}/active?s=t1\&e=t2.  This URI will return all active segments for the given FLAG in the GPS interval [t1, t2). 

The \ac{DQSEGDB} database service with this design is much faster and more stable than its predecessors.  The database currently holds over 600 million DQ segment entries stably. The server responds to queries such as a request for the segments that define when the LIGO Livingston IFO is in optimal observing mode over a month of operations, within 3~s.    The MariaDB database is well optimized, such that most of this time is spent in the conversion to the JSON output format. Table \ref{table1} lists some metrics of the performance of the server in 2016 and 2020.  The database has grown by a factor of O(100) over this time, and the results demonstrate that the performance of the server has not changed despite this increase in size.  The server also responds to O(5) times more requests per day now than in 2016, averaging 7.178 requests per second.

\begin{table}[h!]
\centering
\begin{tabular}{c | c c c c | c c c} 
\toprule
\multirow{2}{*}{Date} & \multicolumn{4}{c@{}}{Requests} & \multicolumn{3}{c@{}}{Avg. Response Time (s)} \\
\cmidrule(l){2-8}
& Get & Patch & Total & Req./s & Get & Patch & All \\
\midrule
09/14/16 & 37,673 & 80,857 & 118,530 & 1.401 & 3.584 & 0.018 & 1.151 \\
02/10/20 & 39,690 & 567,588 & 607,278 & 7.178 & 3.238 & 0.012 & 0.223 \\
\bottomrule
\end{tabular}
\caption{Demonstrating performance stability of the DQSEGDB service.  In 2020, the database contains O(100) times more data, and responds to O(5) times more requests per second with nearly identical performance compared to 2016 values.}
\label{table1}
\end{table}




\section{Impact}
\label{}



The new \ac{DQSEGDB} system of servers has been very successful in meeting the needs of the GW community for storing and distributing \ac{IFO} metadata since 2014.  Thanks to the high performance of the \ac{DQSEGDB} service, nearly all \ac{LVC} GW searches are using this centralized source of data quality information. It is, thus, also providing a system for careful control and synchronization of the detector status information used by the \ac{LVC} searches.  The \ac{DQSEGDB} is also used by many automated \ac{IFO} monitoring processes and many \ac{LVC} scientists investigating the performance of the \ac{IFO}s.  In particular, the data analyses that concluded in the detection of all GWs thus far have relied on the DQSEGDB infrastructure \cite{gw150914,gw170817,2020arXiv200101761T,GBM:2017lvd,Abbott:2017oio}.  

\begin{figure}[h] 
\centerline{\includegraphics[width=0.7\textwidth]{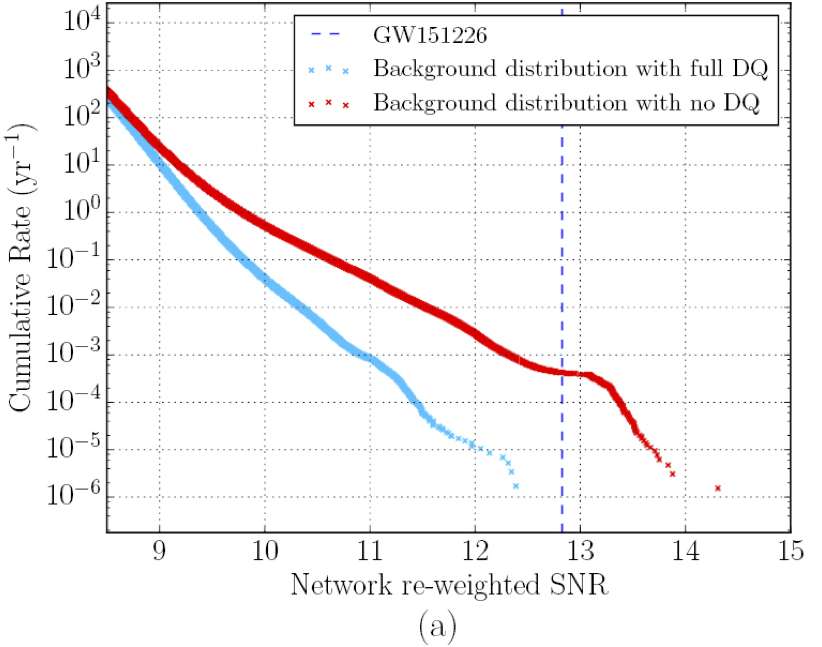}}
\caption{This image is reproduced with permission from \cite{Abbott:2017lwt}, original creator T.~J.~Massinger. This image illustrates the impact of applying \ac{DQ} flags contained in the DQSEGDB to a \ac{LVC} GW search in the data from Advanced LIGO's first observing run.  The initial background events in the search have \ac{SNR} values reaching above 14. With the \ac{DQ} applied, the background is reduced to below an \ac{SNR} of 12.5.  The upper limit of this background is closely tied to the limit at which GW events may confidently be detected.  The \ac{SNR} of the GW event GW151226 is indicated on the figure to demonstrate that this detection would have been missed without the use of the DQ data.  Image License \url{https://creativecommons.org/licenses/by/3.0/}, image was cropped from original.
}
\label{fig:massinger_impact}
\end{figure}

The impact of the \ac{DQ} information hosted in the DQSEGDB on GW searches is significant, as illustrated in Figure \ref{fig:massinger_impact} \cite{Abbott:2017lwt}.  The information is used to mitigate systematic noise issues, and thanks to the speed of the \ac{DQSEGDB} service, the speed of testing of the different choices of \ac{DQ} flags has been drastically improved.  One example of the types of DQ flags used to remove a significant amount of noise was the "RF45 flag".  This flag indicated times when issues with the electronics that controlled the radio frequency (RF) sidebands used to sense and control LIGO’s optical cavities would contaminate the main detection channel with noise that resulted in a significant number of false triggers in \ac{DA} pipelines \cite{Abbott:2017lwt}.  The latency from the time data is collected at the IFO sites to the moment the metadata may be queried by rapid analyses has also been reduced to less than 5 minutes. This functionality is being used by several ``medium latency" analyses that are automatically started in response to external events such as observations of gamma-ray bursts.  

The new DQSEGDB service has allowed new collaboration tools that automatically, and very frequently, query the \ac{DQ} flag metadata to be developed.  These services provide many different benefits to the large, 1000+ person LVC. In particular, the Summary Page web infrastructure \cite{duncan_macleod_2019_3590375} makes heavy use of this ability. One open-access example of these pages is available at \url{https://www.gw-openscience.org/detector_status/day/20200207/}.  Many of the plots indicate the state of the interferometers, and all of this metadata is being retrieved from the DQSEGDB.  These plots are updated on a rolling basis, requiring very frequent queries to the DQSEGDB service that the old service would not have been able to handle.  These pages are used by \ac{IFO} commissioners, data quality investigators, data analysts and the wider astronomical community to easily assess the state of the interferometers and rapidly investigate systematic issues.  They have proven invaluable to data and event validation efforts in addition to daily \ac{IFO} and collaboration operations. 

Finally, due to the speed and reliability of the service, additional GW detectors have also begun using this single instance of the DQSEGDB.  The \ac{GEO} collaboration now uses this DQSEGDB instance to store its primary detector state data.  The \ac{KAGRA} collaboration has recently begun storing metadata with this service as well, with the recent start of full observations.  Thus, the DQSEGDB infrastructure and service is now used by all IFO-based GW detection efforts in the world.






\section{Conclusions}
\label{}
The DQSEGDB software has been tremendously successful in serving the GW astronomy community. This set of database, backend, frontend and client software has provided rapid access to the DQ segments needed by the \ac{LVC} for all GW detections made thus far.  The speed and reliability of the database combined with its clean, RESTful API has resulted in the design of new tools that enable scientists to more rapidly and easily understand the \ac{IFO}s in the GW detection network.  

\section{Conflict of Interest}

No conflict of interest exists:
We wish to confirm that there are no known conflicts of interest associated with this publication and there has been no significant financial support for this work that could have influenced its outcome.

\section{Role of Funding Source}

Funding for this project was provided by the National Science Foundation, the European Gravitational Observatory, the Italian Istituto Nazionale di Fisica Nucleare (INFN),  
the French Centre National de la Recherche Scientifique (CNRS) and
the Foundation for Fundamental Research on Matter supported by the Netherlands Organisation for Scientific Research.  The funding sources had no involvement in the work described in this article or the writing or submission of the article.

\section*{Acknowledgements}
\label{acknowledgements}
The authors gratefully acknowledge the support of the United States
National Science Foundation (NSF) for the construction and operation of the
LIGO Laboratory and Advanced LIGO as well as the Science and Technology Facilities Council (STFC) of the
United Kingdom, the Max-Planck-Society (MPS), and the State of
Niedersachsen/Germany for support of the construction of Advanced LIGO 
and construction and operation of the GEO600 detector.  The authors gratefully acknowledge the National Science Foundation Grants PHY-1700765 and PHY-1104371 for the support of this project.  
Additional support for Advanced LIGO was provided by the Australian Research Council.
The authors gratefully acknowledge the Italian Istituto Nazionale di Fisica Nucleare (INFN),  
the French Centre National de la Recherche Scientifique (CNRS) and
the Foundation for Fundamental Research on Matter supported by the Netherlands Organisation for Scientific Research, 
for the construction and operation of the Virgo detector, their support in all software developments, and the creation and support of the EGO consortium. 
The authors also gratefully acknowledge research support from these agencies as well as by 
the Council of Scientific and Industrial Research of India, 
the Department of Science and Technology, India,
the Science \& Engineering Research Board (SERB), India,
the Ministry of Human Resource Development, India,
the Spanish  Agencia Estatal de Investigaci\'on,
the Vicepresid\`encia i Conselleria d'Innovaci\'o, Recerca i Turisme and the Conselleria d'Educaci\'o i Universitat del Govern de les Illes Balears,
the Conselleria d'Educaci\'o, Investigaci\'o, Cultura i Esport de la Generalitat Valenciana,
the National Science Centre of Poland,
the Swiss National Science Foundation (SNSF),
the Russian Foundation for Basic Research, 
the Russian Science Foundation,
the European Commission,
the European Regional Development Funds (ERDF),
the Royal Society, 
the Scottish Funding Council, 
the Scottish Universities Physics Alliance, 
the Hungarian Scientific Research Fund (OTKA),
the Lyon Institute of Origins (LIO),
the Paris \^{I}le-de-France Region, 
the National Research, Development and Innovation Office Hungary (NKFIH), 
the National Research Foundation of Korea,
Industry Canada and the Province of Ontario through the Ministry of Economic Development and Innovation, 
the Natural Science and Engineering Research Council Canada,
the Canadian Institute for Advanced Research,
the Brazilian Ministry of Science, Technology, Innovations, and Communications,
the International Center for Theoretical Physics South American Institute for Fundamental Research (ICTP-SAIFR), 
the Research Grants Council of Hong Kong,
the National Natural Science Foundation of China (NSFC),
the Leverhulme Trust, 
the Research Corporation, 
the Ministry of Science and Technology (MOST), Taiwan
and
the Kavli Foundation.
The authors gratefully for computational resources provided by the NSF, STFC, INFN and CNRS, and supported by National Science Foundation Grants PHY-1626190, PHY-1700765, PHY-0757058 and PHY-0823459.  
The authors gratefully acknowledge the contributions of Larne Pekowsky and Ping Wei for the development, support and improvement of the original LIGO segment database and Leone Bosi for the development of the original Virgo database. 






\bibliographystyle{elsarticle-num} 
\bibliography{proposalbib}

\section*{Required Metadata}
\label{}

\section*{Current code version}
1.6.1

\begin{table}[H]
\begin{tabular}{|l|p{6.5cm}|p{6.5cm}|}
\hline
\textbf{Nr.} & \textbf{Code metadata description} & \textbf{Please fill in this column} \\
\hline
C1 & Current code version & 1.6.1 \\
\hline
C2 & Permanent link to code/repository used for this code version & \url{https://github.com/ligovirgo/dqsegdb/tree/dqsegdb-release-1.6.1} \\
\hline
C3 & Code Ocean compute capsule & N/A\\
\hline
C4 & Legal Code License   & GNU GENERAL PUBLIC LICENSE v3.0 \\
\hline
C5 & Code versioning system used & git \\
\hline
C6 & Software code languages, tools, and services used & Python, MariaDB, PHP \\
\hline
C7 & Compilation requirements, operating environments \& dependencies & Scientific Linux 7.5, Python 2.7.5 \\
\hline
C8 & If available Link to developer documentation/manual & \url{http://ligovirgo.github.io/dqsegdb/} \\
\hline
C9 & Support email for questions & question@ligo.org \\
\hline
\end{tabular}
\caption{Code metadata (mandatory)}
\label{} 
\end{table}

\end{document}